\documentclass[10pt]{iopart}

\expandafter\let\csname equation*\endcsname\relax

\expandafter\let\csname endequation*\endcsname\relax
\usepackage{graphicx}
\usepackage{hyperref}
\usepackage{xurl}
\usepackage{har2nat}
\usepackage{multirow}
\usepackage{float}
\usepackage{comment}
\usepackage{subcaption}
\usepackage{hhline}
\usepackage{amsmath}
\usepackage{tabularx}
\usepackage{amsthm,amsfonts}
\usepackage[labelsep=colon,
            	labelfont={sf,bf},
                textfont=sf]{caption}
\usepackage[dvipsnames]{xcolor}

\bibpunct[, ]{(}{)}{,}{a}{}{,}%

\begin{document}

\title[Multiscale Partial Information Decomposition Under Long-Range Correlations]{Multiscale Partial Information Decomposition of Dynamic Processes with Short and Long-range correlations: Theory and Application to Cardiovascular Control}

\author{H\'{e}lder Pinto$^{1,2}$, Riccardo Pernice$^3$, Maria Eduarda Silva$^{4,5}$, Michal Javorka$^6$, Luca Faes$^3$, Ana Paula Rocha$^{1,2}$}

\address{$^1$ Faculdade de Ci\^{e}ncias, Universidade do Porto, Portugal}
\address{$^2$ Centro de Matem\'{a}tica da Universidade do Porto (CMUP), Porto, Portugal}
\address{$^3$ Department of Engineering, University of Palermo, Palermo, Italy}
\address{$^4$ Faculdade de Economia, University do Porto, Portugal}
\address{$^5$ LIAAD-INESC TEC, Porto, Portugal}
\address{$^6$ Department of Physiology, Jessenius Faculty of Medicine, Comenius University of Bratislava, Martin, Slovakia}
\ead{helder.pinto@fc.up.pt, riccardo.pernice@unipa.it, mesilva@fep.up.pt, michal.javorka@uniba.sk, luca.faes@unipa.it, aprocha@fc.up.pt}
\vspace{10pt}
\begin{indented}
\item[]March 2022
\end{indented}

\begin{abstract}
Heart rate variability results from the combined activity of several physiological systems, including the cardiac, vascular, and respiratory systems which  have their own internal regulation, but also interact with each other to preserve the homeostatic function. These control mechanisms operate across multiple temporal scales, resulting in the simultaneous presence of short term dynamics and long-range correlations. The Network Physiology framework provides statistical tools based on information theory able to quantify structural aspects of multivariate and multiscale interconnected mechanisms driving the dynamics of complex physiological networks. In this work, the multiscale representation of Transfer Entropy from Systolic Arterial Pressure (S) and Respiration (R) to Heart Period (H) and of its decomposition into unique, redundant and synergistic contributions is obtained using a Vector AutoRegressive Fractionally Integrated (VARFI) framework for Gaussian processes. This novel approach allows to quantify the directed information flow accounting for the simultaneous presence of short-term dynamics and long-range correlations among the analyzed processes. Additionally, it provides analytical expressions for the computation of  the information measures, by exploiting the theory of state space models. The approach is first illustrated in simulated VARFI processes and then applied to H, S and R time series measured in healthy subjects monitored at rest and during mental and postural stress. Our results highlight the dependence of the information transfer on the balance between short-term and long-range correlations in coupled dynamical systems, which cannot be observed using standard methods that do not take into account long-range correlations. The proposed methodology shows that postural stress induces larger redundant effects at short time scales and mental stress induces larger cardiovascular information transfer at longer time scales, thus evidencing the different nature of the two stressors.
\end{abstract}

%
\vspace{2pc}

%
%
\ioptwocol

\section{Introduction} \label{sec:Introduction}

Cardiovascular oscillations result from the activity of several coexisting control mechanisms also interconnected with respiratory activity and, as a consequence, exhibit a complex dynamical structure \citep{cohen2002short}. The action of these simultaneously active and intertwined mechanisms occurs in multiples time scales and is reflected in the spontaneous beat-to-beat variability of the Heart Period (H), and Systolic Arterial Pressure (S) continuously interacting with Respiratory activity (R). The multivariate and simultaneous analysis of cardiovascular oscillations can be very helpful to understand the network of interconnections among variables,  shedding light on the combined activity of physiological mechanisms like the baroreflex and the Respiratory Sinus Arrhythmia (RSA)
\citep{lanfranchi2002arterial,berntson1993respiratory}. Heart rate variability (HRV), i.e. the changes in the time intervals between consecutive heartbeats, differs in relation to the time scale at which processes are observed, corresponding to a given frequency of heart rate oscillations. Coupling and causality between the regulatory processes can be associated to specific time scales of oscillatory activity. Therefore, it is important to characterize the mechanisms and interactions governing heart rate variability on various time scales \citep{faes2004causal,cerutti2009multiscale}. Another important feature of the cardiovascular control mechanisms and cardiorespiratory interactions is the presence of long range correlations, resulting in slowly varying dynamics \citep{martins2020multivariate,faes2019multiscalePRE,xiong2017entropy}.

Cardiovascular and cardiorespiratory interactions are often studied using information-theory applied to the dynamics of the H, S and R time series \citep{faes2017information_book,faes2015information}. In particular, emerging information-theoretic approaches such as the so-called “Partial Information Decomposition” (PID) and the "Interaction Information Decomposition" (IID) allow assessing the information transfer among the multiple nodes of a network system \citep{lizier2018information}.
Such approaches can be contextualized within the general field of “Network Physiology”, which describes the human body as an integrated network where multiple organs continuously interact with each other reflecting various physiological and pathological states \citep{ivanov2021new,bashan2012network}. The IID and PID frameworks have been used to decompose, in a network composed of multiple processes, the information flowing from two sources to a target into unique contributions related to each individual source, and to separate synergistic and redundant contributions.

The present study aims to extend multiscale partial information decomposition \citep{faes2017multiscale} to the combined analysis of short-term and long-range correlations among coupled processes, and to employ the extended approach to quantify the amount of information transferred from S and R towards H, and also to identify the type of interaction (synergistic or redundant) between S and R while they transfer information to H. To this end, we propose a method using vector autoregressive fractionally integrated ($\mathrm{VARFI}$) models which provides the multiscale representation of the $\mathrm{VARFI}$ parameters using the theory of state space models, thereby allowing to extract from such parameters multiscale and multivariate information transfer measures \citep{faes2019multiscalePRE,martins2020multivariate}. The advantages of this method resides in its parametric formulation that permits to work reliably on short time series and in the operation of fractional integration that allows to take into account not only short-term dynamics, but also the long-range correlations. Furthermore, decomposing the information allows assessing the directionality of the interactions, which has not been done in previous works \citep{martins2020multivariate,faes2019multiscalePRE} where information measures like the complexity and the information storage were used.

This approach is first tested in simulations of a benchmark VARFI model and then applied to experimental data consisting of H, S and R time series measured in healthy subjects monitored in a relaxed physiological condition (supine position) and during two types of stress: postural stress provoked by head-up tilt and mental stress induced by mental arithmetic test.

\section{Methods} \label{sec:Methods}

Let us consider a dynamical system $\mathcal{X}$ whose activity is defined by a discrete-time, stationary vector stochastic process composed of $M$ real-valued zero-mean scalar processes $X_{j,n}$ with $j=1 \ldots M$, $\mathbf{X}_{n}=\left[X_{1, n} \cdots X_{M, n}\right]^{T},-\infty<n<\infty.$ The past of  the scalar processes is denoted as $X_{j, n}^{-}=\left[X_{j, n-1}, X_{j, n-2} \cdots\right].$ With this notation, in the following we present our methodology to assess information decomposition in multiple processes accounting for short-term dynamics and long-range correlations.

\subsection{Information Transfer and Modification} \label{sec:IID&PID}

In an information-theoretic framework, the directed transfer of information between scalar sub-processes is assessed by the Transfer Entropy (TE). Specifically, the TE quantifies the amount of information that the past of the source provides about the present of the target process over and above the information already provided by the past of the target itself \citep{schreiber2000measuring}. Transfer entropy between the source process component $i$ and the target component $j$ is defined as

\begin{equation} \label{eq:IndividualTE}
T_{i \rightarrow j}=I\left(X_{j, n} ; X_{i, n}^{-} \mid X_{j, n}^{-}\right),
\end{equation}
where $I(\cdot ; \cdot \mid \cdot)$ denotes conditional Mutual Information (MI) \citep{Cover2005}. Considering two sources $X_{i}$ and $X_{k}$ and a target $X_{j}$, the information transferred towards $X_{j}$ from the sources $X_{i}$ and $X_{k}$ taken together is quantified by the joint TE (JTE)

\begin{equation} \label{eq:JTE}
T_{i k \rightarrow j}=I\left(X_{j, n} ; X_{i, n}^{-}, X_{k, n}^{-} \mid X_{j, n}^{-}\right).
\end{equation}

Generally, the joint TE differs from the sum of the two individual TEs, since the source processes ($i, k$) typically interact with each other while they transfer information to the target process, $j$. The joint TE \eref{eq:JTE} can be decomposed under an Interaction Information Decomposition (IID) framework as \citep{faes2017multiscale}

\begin{equation} \label{eq:IID}
T_{i k \rightarrow j}=T_{i \rightarrow j}+T_{k \rightarrow j}+I_{i k \rightarrow j},
\end{equation}
where $I_{i k \rightarrow j}$ is denoted as Interaction Transfer Entropy (ITE) since it is equivalent to the interaction information \citep{mcgill1954multivariate} computed between the present of the target and the past of the two sources, conditioned on the past of the target

\begin{equation} \label{eq:ITE}
I_{i k \rightarrow j}=I\left(X_{j, n} ; X_{i, n}^{-} ; X_{k, n}^{-} \mid X_{j, n}^{-}\right).
\end{equation}

The ITE quantifies the modification of the information transferred from the source processes $X_i$ and $X_k$ to the target $X_j$. The ITE can take positive and negative values. Positive values of $I_{i k \rightarrow j}$ denote synergy, where the joint TE  is greater than the sum of the two individual TEs $\left(T_{i k \rightarrow j}>T_{i \rightarrow j}+T_{k \rightarrow j}\right)$. In contrast, negative values of $I_{i k \rightarrow j}$ refer to redundancy, which occurs when the information transferred from the sources to the target overlapped, meaning that the sum of individual TEs is larger than the joint TE $\left(T_{i \rightarrow j}+T_{k \rightarrow j}>T_{i k \rightarrow j}\right)$. The main drawback of IID is that the interaction TE is quantified using only one measure, and thus makes redundancy and synergy mutually exclusive. 

This disadvantage can be overcome by the Partial Information Decomposition (PID) \citep{williams2010nonnegative} encompassing four distinct positive quantities \citep{faes2017multiscale} 
\begin{subequations} \label{eq:PID}
\begin{equation} 
T_{i k \rightarrow j}=U_{i \rightarrow j}+U_{k \rightarrow j}+R_{i k \rightarrow j}+S_{i k \rightarrow j},
\end{equation}
\begin{equation}
T_{i \rightarrow j}=U_{i \rightarrow j}+R_{i k \rightarrow j},
\end{equation}
\begin{equation}
T_{k \rightarrow j}=U_{k \rightarrow j}+R_{i k \rightarrow j} .
\end{equation}
\end{subequations}
The terms $U_{i \rightarrow j}$ and $U_{k \rightarrow j}$ quantify the parts of the information transferred to the target process $X_{j}$, which are unique to the source processes $X_{i}$ and $X_{k}$, respectively, thus reflecting contributions to the predictability of the target that can be obtained from one of the sources alone. Then, the terms $R_{i k \rightarrow j}$ and $S_{i k \rightarrow j}$ quantify the redundant and synergistic interaction between the two sources and the target, respectively.

When compared to IID \eref{eq:IID}, the PID \eref{eq:PID}  has the advantage that it provides distinct non-negative measures of redundancy and synergy, therefore allowing the simultaneous presence of redundancy and synergy as distinct elements of information modification. Moreover, the IID and PID are related to each other as
\begin{equation} \label{eq:RelationITE}
I_{i k \rightarrow j}=S_{i k \rightarrow j}-R_{i k \rightarrow j},
\end{equation}
thus showing that the interaction TE is actually a measure of the "net" synergy manifested in the transfer of information from the two sources to the target \citep{krohova2019multiscale}.

The main issue with the PID \eref{eq:PID} is that its constituent measures cannot be obtained from the classic information theory simply subtracting conditional MI terms, as done for the IID. Therefore, to complete the PID an additional ingredient to the theory is needed to get a fourth defining equation to be added to \eref{eq:PID} for providing an unambiguous definition of $U_{i \rightarrow j}, U_{k \rightarrow j}, R_{i k \rightarrow i}$ and $S_{i k \rightarrow j}.$ Several PID definitions have been proposed arising from different conceptual definitions of redundancy and synergy \citep{harder2013bivariate,griffith2014intersection,bertschinger2014quantifying}. Here we make reference to the so-called Minimum Mutual Information PID (MMI-PID) \citep{barrett2015exploration}. In this approach, redundancy is defined as the minimum of the information provided by each individual source to the target. This leads to the following definition of the redundant TE

\begin{equation} \label{eq:MMI PID}
R_{i k \rightarrow j}=\min \left\{T_{i \rightarrow j}, T_{k \rightarrow j}\right\}.
\end{equation}

This definition satisfies the desirable property that the redundant TE is independent of the correlation between the source processes. Furthermore,  if the observed processes have a joint Gaussian distribution, all previously-proposed PID formulations reduce to the MMI PID \citep{barrett2015exploration}.

\subsection{Vector Autoregressive Fractionally Integrated Model} \label{sec:VARFI Model}

The classic parametric approach to describe linear Gaussian stochastic processes exhibiting both short-term dynamics and long-range correlations is based on representing an M-dimensional discrete-time, zero-mean and unitary variance stochastic process $\mathbf{X}_{n}$ as a Vector Autoregressive Fractionally Integrated ($\mathrm{VARFI}$) process fed by uncorrelated Gaussian innovations $\mathbf{E}_{n}$. The $\mathrm{VARFI}(p,\mathbf{d})$ process is expressed as \citep{tsay2010maximum}

\begin{equation} \label{eq:VARFIMODEL}
\mathbf{A}(L)\mathrm{diag}\left(\nabla^{\mathbf{d}}\right)\mathbf{X}_{n}=\mathbf{E}_{n},
\end{equation}
where $L$ is the back-shift operator $\left(L^{i} \mathbf{X}_{n}=\mathbf{X}_{n-i}\right)$, $\mathbf{A}(L)=\mathbf{I}_{M}-\sum_{i=1}^{p} \mathbf{A}_{i} L^{i}$ ($\mathbf{I}_{M}$ is the identity matrix of size $M$) is a vector autoregressive (VAR) polynomial of order $p$ defined by the $M \times M$ coefficient matrices $\mathbf{A}_{1}, \ldots, \mathbf{A}_{p}$, and

\begin{equation*} \label{eq:diagMatrix}
\mathrm{diag}\left(\nabla^{\mathbf{d}}\right)=\left[\begin{array}{cccc}
(1-L)^{d_{1}} & 0 & \ldots & 0 \\
0 & (1-L)^{d_{2}} & \ldots & 0 \\
\vdots & \vdots & & \vdots \\
0 & 0 & \ldots & (1-L)^{d_{M}}
\end{array}\right],
\end{equation*}
and $(1-L)^{d_{i}}, i=1, \ldots, M$, is the fractional differencing operator defined by:

\begin{equation} \label{eq:diffoperator}
(1-L)^{d_{i}}=\sum_{k=0}^{\infty} G_{k}^{(i)} L^{k}, \quad G_{k}^{(i)}=\frac{\Gamma\left(k-d_{i}\right)}{\Gamma\left(-d_{i}\right) \Gamma(k+1)},
\end{equation}
with $\Gamma(\cdot)$ denoting the Gamma (generalized factorial) function. The $\mathrm{VARFI}$ model is stationary when all the roots of $\operatorname{det}[\boldsymbol{A}(L)]$ are outside the unit circle and $-0.5 < d_i < 0.5$ for $i=1 \ldots M,$ while it is nonstationary but mean reverting for $0.5 \leq d_i < 1$ \citep{velasco1999gaussian, baillie1996long}. The coefficients of the polynomial $\mathbf{A}(L)$ allow the description of the short term dynamics, while the parameter $\mathbf{d}=\left(d_{1}, \ldots, d_{M}\right)$ in \Eref{eq:VARFIMODEL} determines the long-term behavior of each individual process.

The parameters of the $\mathrm{VARFI}(p,\textbf{d})$ model \eref{eq:VARFIMODEL}, namely the coefficients of $\mathbf{A}(L)$ and the variance of the innovations $\Sigma_{\mathbf{E}}=\mathbb{E}\left[\mathbf{E}_{n}^{T} \mathbf{E}_{n}\right]$, are generally obtained from process realizations of finite length first estimating the differencing parameters $d_i$ using the Whittle semi-parametric local estimator \citep{beran2016long} individually for each process $X_i$; then defining the filtered data $X_{i, n}^{(f)}=(1-L)^{d_{i}} X_{i, n}$; and finally estimating the VAR parameters from the filtered data $\mathbf{X}_{n}^{(f)}$ using the ordinary least squares method to solve the VAR model $\mathbf{A}(L) \mathbf{X}_{n}^{(f)}=\mathbf{E}_{n}$, with model order $p$ assessed through the Bayesian information criterion \citep{faes2012measuring}.

Here, we approximate the VARFI process \eref{eq:VARFIMODEL} with a finite order VAR process by truncating the fractional integration at a finite lag $q$, as follows

\begin{equation} \label{eq:DiffTrunc}
\begin{split}
& \mathrm{diag}\left(\nabla^{\mathrm{d}}\right) \approx \mathbf{G}(L)
\\&=\mathrm{diag}\left[\sum_{k=0}^{q} G_{k}^{(1)} L^{k} \quad \ldots \quad \sum_{k=0}^{q}  G_{k}^{(M)} L^{k}\right] \\
&=\sum_{k=0}^{q} \mathrm{diag}\left[G_{k}^{(1)}, \ldots, G_{k}^{(M)}\right] L^{k} 
\\
&=\sum_{k=0}^{q} \mathbf{G}_{k} L^{k}.
\end{split}
\end{equation}
This allows us to express the $\mathrm{VARFI}(p, \mathbf{d})$ process as a $\mathrm{VAR}(m)$ process, with $m=p+q$
\begin{equation} \label{eq:VAR(m)}
\mathbf{B}(L) \mathbf{X}_{n}=\mathbf{E}_{n},
\end{equation}
with the coefficients in $\mathbf{B}(L)$ given by
\begin{equation} \label{eq:NewCoeffExpansion}
\begin{split}
\mathbf{B}(L)&=\mathbf{A}(L) \mathbf{G}(L)
\\
&=\left(\mathbf{I}_{M}-\sum_{i=1}^{p} \mathbf{A}_{i} L^{i}\right)\left(\sum_{k=0}^{q} \mathbf{G}_{k} L^{k}\right)
\\
&=\mathbf{I}_{M}-\sum_{k=0}^{p+q} \mathbf{B}_{k} L^{k},
\end{split}
\end{equation}
yielding, for $q \geq p,$
\begin{equation}\label{eq:NewCoeff}
\begin{split}
\textbf{B}_0&=\textbf{I}_M \quad , \\
\textbf{B}_k&=\begin{cases}
-\textbf{G}_k+\sum\limits_{i=1}^k \textbf{A}_{i}\textbf{G}_{k-i} ,\quad k=1,\ldots,p 	\\
-\textbf{G}_k+\sum\limits_{i=1}^p \textbf{A}_{i}\textbf{G}_{k-i} ,\quad k=p+1,\ldots,q  \\
\sum\limits_{i=0}^{p+q-k} \textbf{A}_{i+k-q}\textbf{G}_{q-i} ,\quad k=q+1,\ldots,q+p  \\
\end{cases}
\end{split} .
\end{equation}

\subsection{Multiscale Representation of VARFI Processes} \label{sec:Multiscale VARFI}

In this section, we describe how to compute the information measures defined in \Sref{sec:IID&PID} across multiple temporal scales, under the hypothesis that the analyzed multivariate process is appropriately modeled by the VARFI representation provided in \Sref{sec:VARFI Model}. The procedure for multiscale analysis extends the rescaling approach proposed in \citep{faes2017multiscale}. Here only the fundamental steps are presented, the mathematical details are provided in \citep{faes2017multiscale} and in the Appendixes of \citep{martins2020multivariate}. 

Typically, to represent a scalar stochastic process at the temporal scale defined by the scale factor $\tau$, a two-step procedure is employed which consists in first filtering the process with a low pass filter with cutoff frequency $f_{\tau}=1 /(2 \tau)$, and then downsampling the filtered process using a decimation factor $\tau$ (taking one every $\tau$ samples) \citep{porta2006complexity,faes2017multiscale}. Extending this approach to the multivariate case, we first implement the following linear finite impulse response (FIR) filter

\begin{equation} \label{VARMA Process}
\mathbf{X}_{n}^{(r)}=\mathbf{D}(L) \mathbf{X}_{n},
\end{equation}
where $r$ denotes the filter order $\mathbf{D}(L)=\sum_{k=0}^{r} \mathbf{I}_{M} D_{k} L^{k}$, and the coefficients of the polynomial $D_{k}, k=1, \ldots, r$, are the same for all scalar processes $X_{j} \in \mathbf{X}$ and are chosen to set up a low pass FIR configuration with cutoff frequency $1 /(2 \tau)$.This step transforms the $\mathrm{VAR}(p+q)$ process \eref{eq:VAR(m)} into a $\mathrm{VARMA}(p+q,r)$ process with moving average (MA) part determined by the FIR filter coefficients

\begin{equation} \label{VARtoVARMA}
\mathbf{B}(L) \mathbf{X}_{n}^{(r)}=\mathbf{D}(L) \mathbf{B}(L) \mathbf{X}_{n}=\mathbf{D}(L) \mathbf{E}_{n}.
\end{equation}

Then, we exploit the connection between $\mathrm{VARMA}$ processes and state space (SS) processes \citep{aoki1991state} to evidence that the VARMA process (10) can be expressed in SS form as

\begin{subequations} \label{VARMAtoSS}
\begin{equation}
\mathbf{Z}_{n+1}^{(r)}=\mathbf{B}^{(r)} \mathbf{Z}_{n}^{(r)}+\mathbf{K}^{(r)} \mathbf{E}_{n}^{(r)},
\end{equation}
\begin{equation}
\mathbf{X}_{n}^{(r)}=\mathbf{C}^{(r)} \mathbf{Z}_{n}^{(r)}+\mathbf{E}_{n}^{(r)},
\end{equation}
\end{subequations}
where $\mathbf{Z}_{n}^{(r)}=\left[\mathbf{X}_{n-1}^{(r)} \cdots \mathbf{X}_{n-m}^{(r)} \mathbf{E}_{n-1} \cdots \mathbf{E}_{n-r}\right]^{T}$ is a $(m+r)$ - dimensional state process, $\mathbf{E}_{n}^{(r)}=\mathbf{D}_{0} \mathbf{E}_{n}$ is the SS innovation process, and the vectors $\mathbf{K}^{(r)}$ and $\mathbf{C}^{(r)}$ and the matrix $\mathbf{B}^{(r)}$ can be obtained from $\mathbf{B}(L)$ and $\mathbf{D}(L).$ Further details can be found in Appendix B of \citep{martins2020multivariate}.

In the second step of the rescaling procedure, the filtered process is downsampled in order to complete the multiscale representation. This is achieved by applying the results in, \citep{faes2017efficient,solo2016state,barnett2015granger} which allow describing the filtered SS process after downsampling in the form

\begin{subequations} \label{DownSampled}
\begin{equation} \label{eq:Down1}
\mathbf{Z}_{n+1}^{(\tau)}=\mathbf{B}^{(\tau)} \mathbf{Z}_{n}^{(\tau)}+\mathbf{K}^{(\tau)} \mathbf{E}_{n}^{(\tau)},
\end{equation}
\begin{equation} \label{eq:Down2}
\mathbf{X}_{n}^{(\tau)}=\mathbf{C}^{(\tau)} \mathbf{Z}_{n}^{(\tau)}+\mathbf{E}_{n}^{(\tau)}.
\end{equation}
\end{subequations}
Equations (\ref{DownSampled}) provide the SS form of the filtered and downsampled version of the original $\mathrm{VARFI}(p,\mathbf{d})$ process, and parameters $\left(\mathbf{B}^{(\tau)}, \mathbf{C}^{(\tau)}, \mathbf{K}^{(\tau)}, \Sigma_{\mathbf{E}^{(\tau)}}\right)$ can be obtained from the SS parameters before downsampling and from the downsampling factor $\tau.$ 

\subsection{Multiscale Information Transfer and Modification} \label{sec:MultiscaleIDD&PID}

In this section, we show how to compute analytically the information decomposition of a jointly Gaussian multivariate stochastic process starting from its associated SS model \eref{DownSampled}.

The derivations are based on the knowledge that the linear parametric representation of Gaussian processes captures all the entropy differences that define the various information measures \citep{barrett2010multivariate}. These entropy differences are related to the partial variances of the present of the target conditioned to its past and the past of one or more sources. The partial variances can be formulated as variances of the prediction errors resulting from linear regression \citep{faes2015information,faes2017information}. Specifically, let us denote as $E_{j \mid j, n}=X_{j, n}-\mathbb{E}\left[X_{j, n} \mid X_{j, n}^{-}\right]$ and $E_{j \mid i j, n}=X_{j, n}-\mathbb{E}\left[X_{j, n} \mid X_{i, n}^{-}, X_{j, n}^{-}\right]$ the prediction errors of a linear regression of $X_{j, n}$ on $X_{j, n}^{-}$ and $\left(X_{j, n}^{-}, X_{i, n}^{-}\right),$ respectively. Then, the TE from $X_{i}$ to $X_{j}$ can be expressed as

\begin{equation} \label{eq:IndividualTE1}
T_{i \rightarrow j}=\frac{1}{2} \ln \frac{\mathbf{\Sigma}_{E_{j \mid j}}}{\mathbf{\Sigma}_{E_{j \mid i j}}}.
\end{equation}

Similarly, the joint TE from $\left(X_{i}, X_{k}\right)$ to $X_{j}$ can be defined as

\begin{equation} \label{eq:JointTE}
T_{i k \rightarrow j}=\frac{1}{2} \ln \frac{\mathbf{\Sigma}_{E_{j \mid j}}}{\mathbf{\Sigma}_{E_{j}}},
\end{equation}
where $\mathbf{\Sigma}_{E_{j}}=\mathbb{E}\left[E_{j,n}^{2}\right]$ is the variance of the prediction error of a linear regression of $X_{j, n}$ on $\mathbf{X}_{n}^{-}, E_{j,n}=X_{j, n}-\mathbb{E}\left[X_{j, n} \mid \mathbf{X}_{n}^{-} \right].$ Based on these derivations, one can easily
complete the IID decomposition of TE by computing $T_{k \rightarrow j}$ as in \eref{eq:IndividualTE1} and deriving the interaction TE from \eref{eq:IID} and the PID decomposition, as well by deriving the redundant TE from \eref{eq:MMI PID}, the synergistic TE from \eref{eq:RelationITE} and the unique TEs from \eref{eq:PID}.

Finally, we show how to compute any partial variance from the parameters of an SS model in the form \eref{DownSampled} at any assigned time scale $\tau$ \citep{solo2016state,barnett2015granger}. The partial variance $\Sigma_{E_{j \mid a}^{(\tau)}}$, where the subscript $a$ denotes any combination of indexes $\in\{1, \ldots, M\}$, can be derived from the SS representation of the innovations of a submodel obtained removing the variables not indexed by $a$ from the observation equation. Specifically, we need to consider the submodel with state \Eref{eq:Down1} and observation equation

\begin{equation} \label{eq:ISSRestr}
\mathbf{X}_{a, n}^{(\tau)}=\mathbf{C}_{a}^{(\tau)} \mathbf{Z}_{n}^{(\tau)}+\mathbf{E}_{a, n}^{(\tau)},
\end{equation}
where the additional subscript $_{a}$ denotes the selection of the rows with indices $a$ in a vector or a matrix. These submodels can be converted to the $\mathrm{SS}$ form as in \eref{DownSampled}, with innovation covariance $\Sigma_{\mathbf{E}_{a}^{(\tau)}}$, so that the partial variance $\Sigma_{E_{j \mid a}^{(\tau)}}$ is derived as the diagonal element of $\Sigma_{\mathbf{E}_{a}^{(\tau)}}$ corresponding to the position of the target $X_{j, n}$ \citep{faes2017multiscale,martins2020multivariate}.

\section{Simulation Study} \label{sec:Simulation}

\begin{figure}[htb!]
\centering
\includegraphics[scale=0.35]{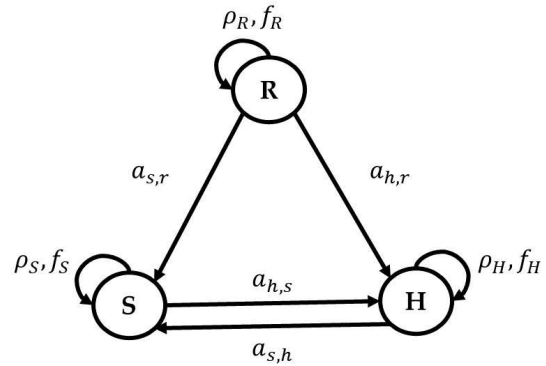}
\caption{Graphical representation of the trivariate VAR process of Equation \ref{eq:SimulatedVAR} with parameters set according to reproduce realistic cardiovascular cardiorespiratory dynamics together with cardiorespiratory interactions.
}
\label{fig:SimulationGraph}
\end{figure}

In this section, we investigate the behavior of the information measures in the presence of long term correlations and considering oscillations and interactions commonly observed in cardiovascular and cardiorespiratory
variability. We start with a $\mathrm{VAR}$ process with short term dynamics described by the benchmark model \citep{faes2017information}:

\begin{equation} \label{eq:SimulatedVAR}
\begin{aligned}
&R_{n}=2 \rho_{r} \cdot \cos 2 \pi f_{r} \cdot R_{n-1}-\rho_{r}^{2} \cdot R_{n-2}+E_{r, n}\\
&S_{n}=2 \rho_{s} \cdot \cos 2 \pi f_{s} \cdot S_{n-1}-\rho_{s}^{2} \cdot S_{n-2}
\\&+a_{s,h} \cdot H_{n-2}+a_{s,r} \cdot R_{n-1}+E_{s, n}\\
&H_{n}=2 \rho_{h} \cdot \cos 2 \pi f_{h} \cdot H_{n-1}-\rho_{h}^{2} \cdot H_{n-2}
\\&+a_{h,s} \cdot S_{n-1}+a_{h,r} \cdot R_{n-1}+E_{h, n}
\end{aligned}
\end{equation}
where $\mathbf{E}_{n}=\left[E_{r, n}, E_{s, n}, E_{h, n}\right]$ is a vector of zero mean white Gaussian noises of unit variance and uncorrelated with each other ($\mathbf{\Sigma}_E=\mathbf{I}$). We set the parameters to reproduce
oscillations and interactions commonly observed in cardiovascular variability in context of cardiorespiratory
interactions, \Fref{fig:SimulationGraph} \citep{malliani1991cardiovascular}, i.e, the self-sustained dynamics typical of Respiration $\left(\mathrm{R}, \rho_{r}=0.9, f_{r}=0.25\right)$ and the slower oscillatory activity commonly observed in the so-called low frequency band in the variability of Systolic Arterial Pressure $\left(\mathrm{S}, \rho_{s}=0.8, f_{s}= 0.1\right)$ and Heart Period $\left(\mathrm{H}, \rho_{h}=0.8, f_{h}=0.1\right)$. The remaining parameters identify causal interactions between processes, which are set from $R$ to $S$ and from $R$ to $H$ (both modulated by the parameter $a_{s,r}=a_{h,r}=1$ ) to simulate the well-known respiration-related fluctuations of arterial pressure and heart rate, and along the two directions of the closed loop between $S$ and $H (a_{s,h}=0.1, a_{h,s}=0.4)$ to simulate bidirectional cardiovascular interactions. All these parameters, summarized in \Tref{tab:Parameters}, were chosen to mimic the oscillatory spectral properties commonly encountered in short-term cardiovascular variability considering the cardiorespiratory interactions.

Next, we provide an approximated simulation of a $\mathrm{VARFI}$ process by using the truncation method \eref{eq:DiffTrunc} introduced in \Sref{sec:Multiscale VARFI}, whereby the vector $\mathbf{d}=\left(d_{r}, d_s, d_{h}\right)$ is provided to assess fractional integration of the three original processes.
With this approach, two simulations are carried out: in Simulation 1 the long range parameters of both $R$ and $H$ processes are kept fixed, while the parameter $d_s$ is increased from 0 to 0.7 (20 points equally spaced on this interval). Then, in Simulation 2 the long range correlations  $\mathbf{d}$ components of the $R$ and $S$ are fixed and the component of $H$ process is increased from 0 to 0.7. In both simulation experiments, we consider the $H$ process as target and the remaining processes $R$ and $S$ as sources. Therefore, in the first experiment we vary the long memory parameter of only one source (in this case $S$) and in the second we fixed the $d$ parameters of the sources and increased the long memory of the target process.

\begin{figure*}[ptbh]
\centering
\begin{subfigure}[b]{0.9\textwidth}
   \centering
   \includegraphics[scale=0.37]{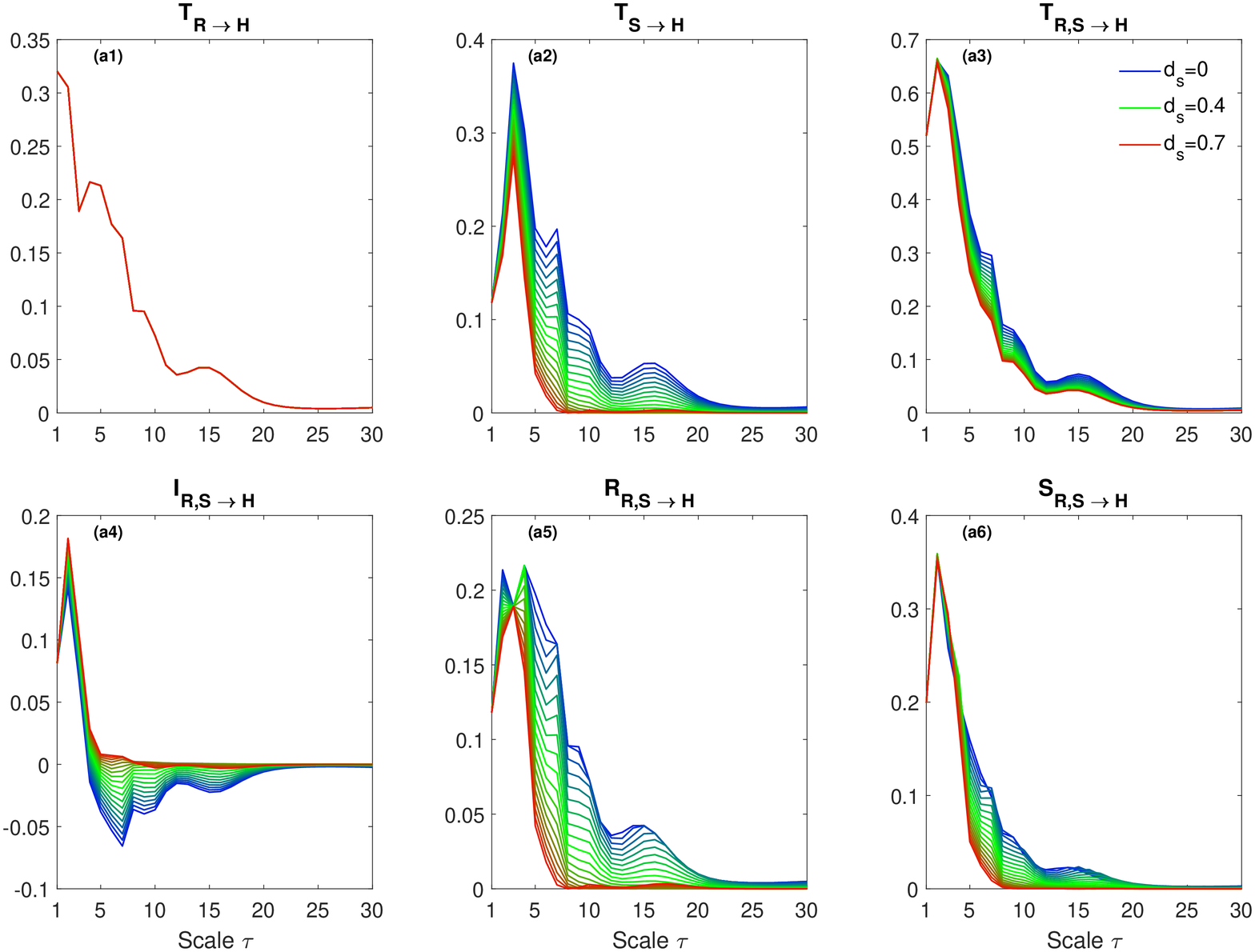}
   \caption{Vary the long range parameter $d_s$ of the $S$ process (source).}
   \label{fig:Simulation_1} 
\end{subfigure}

\begin{subfigure}[b]{0.9\textwidth}
	\centering
   \includegraphics[scale=0.37]{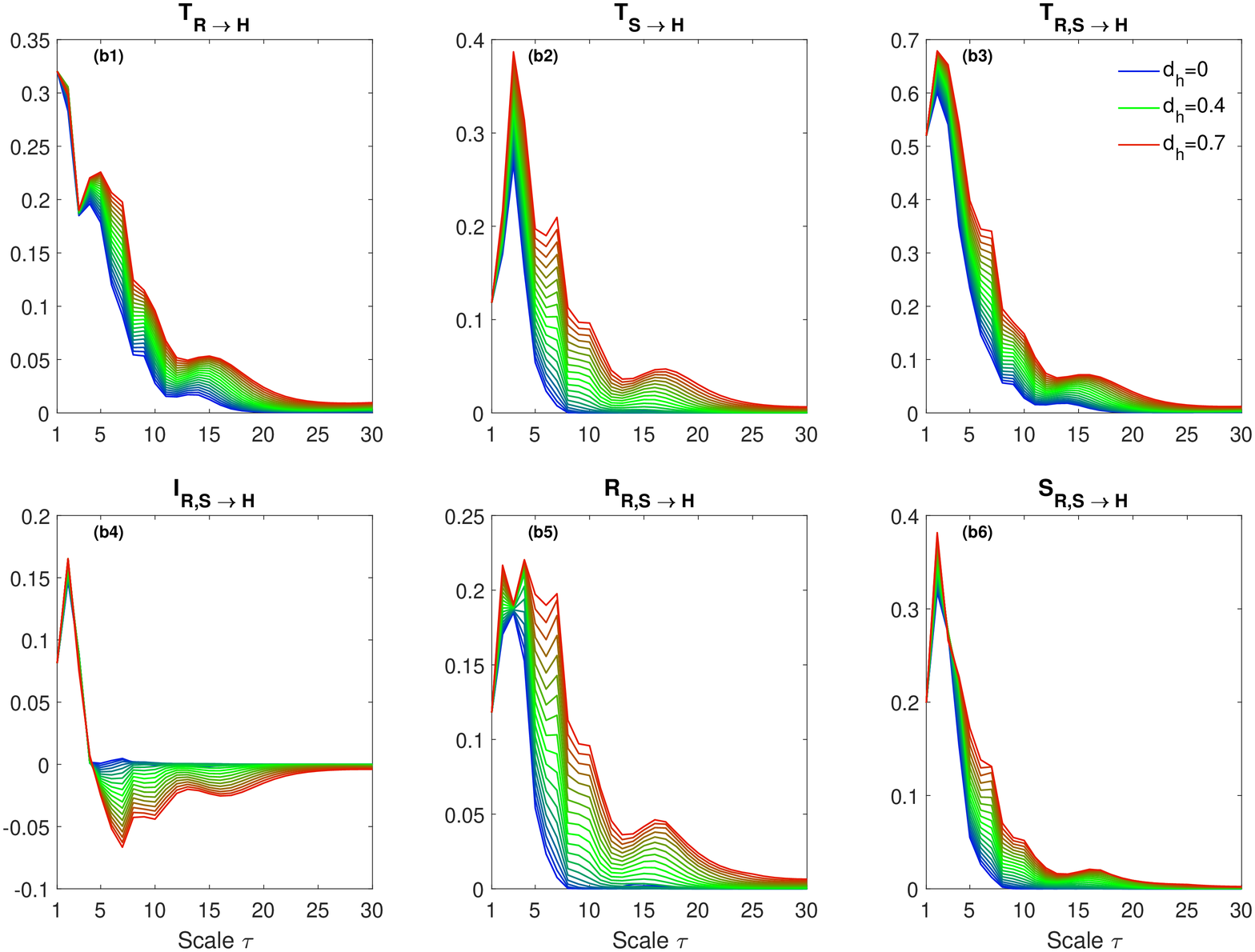}
   \caption{Vary the long range parameter $d_h$ of the $H$ process (target).}
   \label{fig:Simulation_2}
\end{subfigure}

\caption{Theoretical profiles of $T_{S \rightarrow H}, T_{R \rightarrow H}, T_{R, S \rightarrow H}$, $I_{R, S \rightarrow H}$, $R_{R, S \rightarrow H}$ and $S_{R, S \rightarrow H}$ for a $\mathrm{VARFI}$ process. 
\textbf{(a)} Simulation 1: the long memory parameters $d_{r}=0.1$, $d_{h}=0.45$ (source and target) were fixed and $d_{s},$ varies from 0 (blue) to $0.7$ (red).  Note that for $T_{R \rightarrow H}$ all the profiles coincide, as the long range correlation parameters of $R$ and $H$ are the same for all combinations of the $\mathbf{d}$ vectors simulated.\textbf{(b)} Simulation 2: the long memory parameters of the source processes were fixed $d_{r}=0.1$, $d_{s}=0.25$ while the target parameter increased from 0 (blue) to 0.7 (red).}
\label{fig:SimulationsFigs}
\end{figure*}

Figure \ref{fig:SimulationsFigs} reports the results of the individual TEs \eref{eq:IndividualTE}, the joint TE \eref{eq:JTE}, the interaction TE \eref{eq:ITE}, the redundant TE \eref{eq:MMI PID} and the synergistic TE \eref{eq:RelationITE} computed for the overall $\mathrm{VARFI}$ process.

In Figure \ref{fig:SimulationsFigs}(a), when $d_s$ increases, we observe a decrease of the individual TE from $S$ to $H$ and of the joint TE from $R,S$ to $H$ (panels a2 and a3). On the other hand, the ITE increases, suggesting an augmented synergy compared to redundancy. This behavior is better observed when we look at redundancy and synergy separately: comparing the values of these two measures it is visible that as $d_s$ increases synergy is prevalent in relation to redundancy as the latter takes lower values, particularly in the first 10 time scales (panels a4, a5 and a6).

When we vary the long range parameter of the target process, in this case $H$, we observe opposite trends, as seen in Figure \ref{fig:SimulationsFigs}(b): the individual and joint information transfer at longer time scales increase with $d_h$ (panels b1, b2 and b3), and the ITE decreases denoting an increased redundancy. As before, this behavior is clearer when we analyze redundancy and synergy as two distinct elements of information: as $d_h$ increases, redundancy assumes higher values when compared to synergy, particularly on longer time scales, panels b4, b5 and b6.

In summary, our simulations show that the presence of long-range correlations in the source process decreases the information transfer and increases the prevalence of synergy over redundancy, while the opposite behavior (i.e. higher transfer of information and higher redundancy) occurs when long-range correlations are manifested in the target process. These tendencies can explain the role played by long-memory on the transfer of information in complex networks.

\begin{table}[t!]
\centering
\setlength{\tabcolsep}{10pt}
\begin{tabular}{c|c|c|c|c|c|c|c|c|}
\cline{2-9}
\textbf{}                              & \multicolumn{6}{c||}{\textbf{Coupling $a_{u,v}$}}                            & \multicolumn{2}{c|}{\textbf{Poles}} \\ \hhline{|=|=|=|=|=|=|=|=|=|}
\multicolumn{1}{|c|}{\textbf{Process}} & \multicolumn{2}{c|}{R} & \multicolumn{2}{c|}{S}   & \multicolumn{2}{c||}{H}   & $\rho$              & $f$                \\ \hline
\multicolumn{1}{|c|}{R}                & \multicolumn{2}{c|}{-} & \multicolumn{2}{c|}{-}   & \multicolumn{2}{c||}{-}   & 0.9              & 0.25             \\ \hline
\multicolumn{1}{|c|}{S}                & \multicolumn{2}{c|}{1} & \multicolumn{2}{c|}{-}   & \multicolumn{2}{c||}{0.1} & 0.8              & 0.1              \\ \hline
\multicolumn{1}{|c|}{H}                & \multicolumn{2}{c|}{1} & \multicolumn{2}{c|}{0.4} & \multicolumn{2}{c||}{-}   & 0.8              & 0.1              \\ \hline
\end{tabular}
\caption{Parameters of VAR model \eref{eq:SimulatedVAR} that determine the short term dynamics of the VARFI process.}
\label{tab:Parameters}
\end{table}

\section{Cardiovascular Signals Analysis} \label{sec:CRAnalysis}

In this section we apply the proposed approach on experimental data, computing the information measures on cardiovascular and respiratory time series: the heart period ($\mathrm{H}$), systolic arterial pressure ($\mathrm{S}$), and respiration ($\mathrm{R}$). The interaction between the dynamics of these series has been the subject of intense study \citep{krohova2019multiscale,faes2011information,porta2012model,faes2011informationcouplings,faes2004causal}, which motivates their use in a multivariate context. Recent studies have pointed out the intertwined nature of the measures of information dynamics, and the requirement to combine their evaluation to circumvent misinterpretations about the intrinsic network properties \citep{chicharro2012framework,faes2015information,porta2016effect}. In addition, the specificity of measures of information storage and transfer is frequently limited by the fact that their definition incorporates multiple aspects of the dynamical structure of network processes; the flexibility of information measures allows to overwhelm this limitation by decomposing these measures into meaningful quantities \citep{porta2015disentangling}. The most studied variable in cardiovascular spontaneous variability is HRV \citep{malik1996heart,shaffer2017overview,pernice2019comparison}. This variable reflects cardiovascular complexity, strongly interacts with S and R and represents the capability of the organism to react to environmental and psychological stimuli.  For this reason, the target process of the Transfer Entropy measures introduced in \Sref{sec:IID&PID} will be the H time series \citep{krohova2019multiscale,faes2011information,porta2012model,faes2011informationcouplings,faes2004causal}. The assumptions of stationarity and joint Gaussianity that underlie the methodologies presented in this paper are largely exploited in this multivariate analysis, and are usually assumed to hold when realizations of the cardiac, vascular and respiratory processes are obtained in well-controlled experimental protocols designed to achieve stable physiological and experimental conditions \citep{baselli1994model,cohen2002short,faes2012measuring,patton1996baroreflex,triedman1995respiratory,xiao2005system}.

\subsection{Experimental Protocol} \label{sec:ExperimentalProt}

The $\mathrm{H}$, $\mathrm{S}$ and $\mathrm{R}$ time series were measured in a group of 61 healthy subjects (19.5 ± 3.3 years old, 37 females) monitored in the resting supine position ($\mathrm{SU}_1$), in the upright position ($\mathrm{UP}$) reached through passive head-up tilt, in the recovery supine position ($\mathrm{SU}_2$) and during mental stress induced by mental arithmetic test ($\mathrm{MA}$) \citep{javorka2017causal}. We interpreted the previous time series as realizations of the stochastic processes descriptive of the cardiac, vascular, and respiratory dynamics, respectively. For each subject and condition, the analyzed multivariate process is defined as $\mathbf{X}=\left[X_{\mathrm{H}}, X_{\mathrm{S}}, X_{\mathrm{R}}\right]$. During the measurements, the subjects were free-breathing. The acquired signals were the surface electrocardiogram (ECG, horizontal bipolar thoracic lead; CardioFax ECG-9620, NihonKohden, Japan), the finger arterial blood pressure (Finometer Pro, FMS, Netherlands) recorded noninvasively by the volume-clamp photoplethysmographic method, and the respiration signal recorded through respiratory inductive plethysmography (RespiTrace, NIMS, USA). All measured signals were digitised at 1000 Hz. For each subject and experimental condition, the values of H, S and R were measured on a beat-to-beat basis respectively as the sequences of the temporal distances between consecutive R peaks of the ECG, the maximum values of the arterial pressure waveform taken within the consecutively detected heart periods, and the values of the respiratory signal sampled at the onset of the consecutively detected heart periods. The study was approved by Ethical Committee of the Jessenius Faculty of Medicine, Comenius University (Slovakia) and all participants signed a written informed consent. More details about the experimental protocol and signal measurement are reported in \citep{javorka2017causal}.

The analysis was performed on segments of at least 400 consecutive points, free of artifacts and deemed as weak-sense stationary through visual inspection, extracted from the time series for each subject and condition. Missing values and outliers were corrected through linear interpolation and, for H and when possible, erroneous/missing intervals were substituted by pulse intervals measured as the difference in time between two consecutive S measurements $\left(\Delta_{t_{\mathrm{S}}}(n)=t_{\mathrm{S}}(n+1)-t_{\mathrm{S}}(n)\right)$. The three time series were normalized to zero mean and unit variance before multiscale analysis.

\subsection{Data Analysis}

To compute the IID and PID measures, the approach based on complete $\mathrm{VARFI}$ model identification defined in \Sref{sec:Multiscale VARFI} is applied. The $\mathrm{VARFI}$ model is identified first estimating the fractional differencing parameter $d_{i}, i=1, \ldots, 3$, individually for each time series using the Whittle estimator, then filtering the time series with the fractional integration polynomial truncated at a lag $q$, and finally estimating the parameters of the polynomial relevant to the short-term dynamics via least squares $\mathrm{VAR}$ identification. Theoretically, the $\mathrm{VARFI}$ is of infinite order, hence the value of $q$ has to be selected to approximate the $\mathrm{VARFI}$ process with a finite order VAR process. Several previous studies \citep{faes2019multiscalePRE,bardet2003generators} define $q=50$ as an appropriate value for truncating the $\mathrm{VARFI}$ process. By increasing $q$, we can obtain a more precise approximation of the fractional integration part but with a higher computational cost, while a reduced value (and thus an excessive truncation) causes an underestimation of the TE measures and the smoothing of the non-monotonic trends with the time scale \citep{faes2019multiscalePRE}. The order $p$ of the $\mathrm{VAR}$ model was assessed by the Bayesian information
criterion (BIC) \citep{martins2020multivariate}. Then, multiscale TE measures were computed implementing FIR low pass filter of order $r=48$, for time scales $\tau$ in the range $(1, \ldots, 12)$, which corresponds to low pass cutoff normalized frequencies $f_{\tau}=(0.5, \ldots, 0.04)$. The value $r=48$ was  set according to previous settings \citep{faes2019multiscalePRE}.

The differencing parameters $d_{i}$ were estimated individually for each time series in the interval $[-0.5,1[.$ For 3 individuals the estimated $d_{i}$ parameters were near to 1 which indicate that the estimated $\mathrm{VARFI}$ models were nonstationary and thus only 59 subjects were considered for further statistical analysis.

\subsection{Statistical Analysis} \label{sec: Statistical Analysis}

Significant changes in the information transfer and modification measures across the pairs of experimental conditions $\mathrm{SU}_{1}$vs.$\mathrm{UP}$ and $\mathrm{SU}_{2}$vs.$\mathrm{MA}$ are evaluated via a linear mixed-effects model, incorporating both fixed and random effects \citep{pinheiro2006mixed}. The fixed-effects (or factors) were condition and scale, while the random-effect was the subject-dependent intercept that allows for the random variation between subjects. Furthermore, the interaction between the factors is also considered. To assess the changes of interest, estimated marginal means (EMM) \citep{searle1980population} are obtained for each difference, $\mathrm{UP}-\mathrm{SU}_{1}$ and $\mathrm{MA}-\mathrm{SU}_{2}$, at each time scale, $\tau=1, \ldots,12.$ A Z-test is applied to check the significance of these differences at a level $p<0.05$ with the Tukey correction for multiple comparisons. 

Additionally, we used a similar approach to evaluate significant differences between measures, particularly, between the individuals TEs from $S$ to $H$ and from $R$ to $H$, and between the redundant and synergistic TEs, at each time scale $\tau.$ This allows to ascertain which of the source processes, $S$ or $R$, is prevalent in driving $H$ at a given time scale, or whether redundant effects are prevalent over synergistic ones. The comparison between redundancy and synergy is also very important as it provides clues about the type of interactions between source processes, making it possible to corroborate some physiological assumptions. In this case, the fixed-effects (or factors) are measure and scale, and the random-effect is the subject-dependent intercept as in the previous case. For both models, residuals were checked for whiteness. The packages \texttt{lme4} \citep{JSSv067i01} and \texttt{emmeans} \citep{emmeans} of the \texttt{R} software \citep{R} were used to build the models and to compute EMM, respectively.

\section{Results} \label{sec:Results}

This section presents the results of multiscale analysis performed for the IID terms as defined in Equations \eref{eq:IndividualTE}-\eref{eq:ITE}, as well for the redundant and synergistic TEs of the PID Equations \eref{eq:PID} ($R_{i k \rightarrow j}$ and $S_{i k \rightarrow j}$).

\begin{figure*}[ptbh]
\centering
\begin{subfigure}[b]{0.9\textwidth}
   \centering
   \includegraphics[scale=0.4]{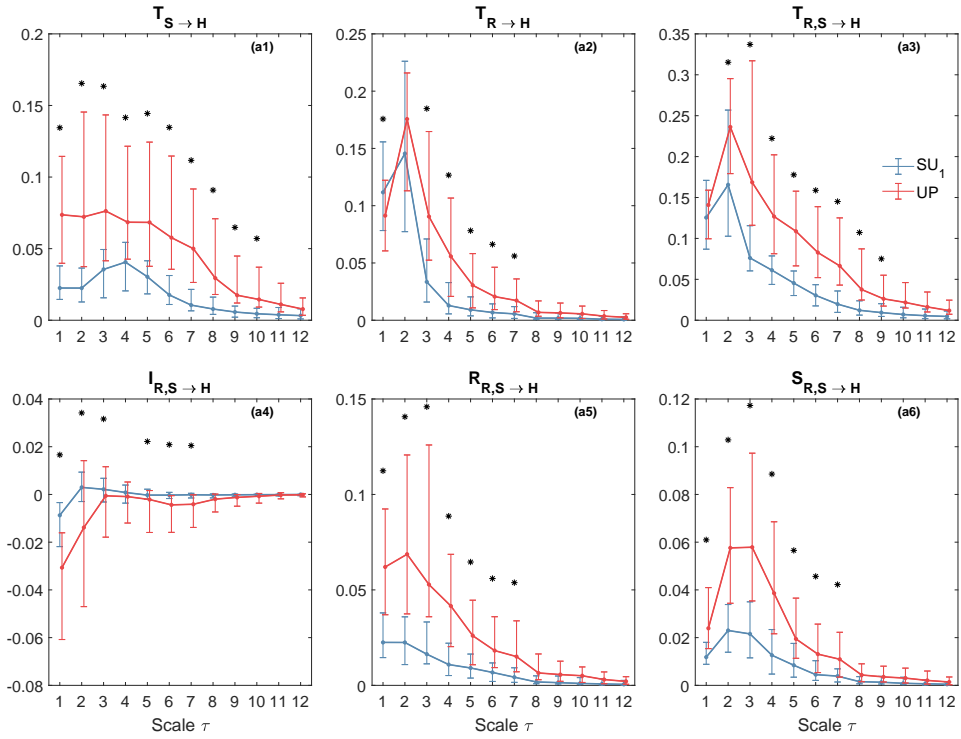}
   \caption{Resting Supine Position $(\mathrm{SU}_1)$ \textit{versus} Upright Position $(\mathrm{UP})$.}
   \label{fig:Analysis_a} 
\end{subfigure}

\begin{subfigure}[b]{0.9\textwidth}
	\centering
   \includegraphics[scale=0.4]{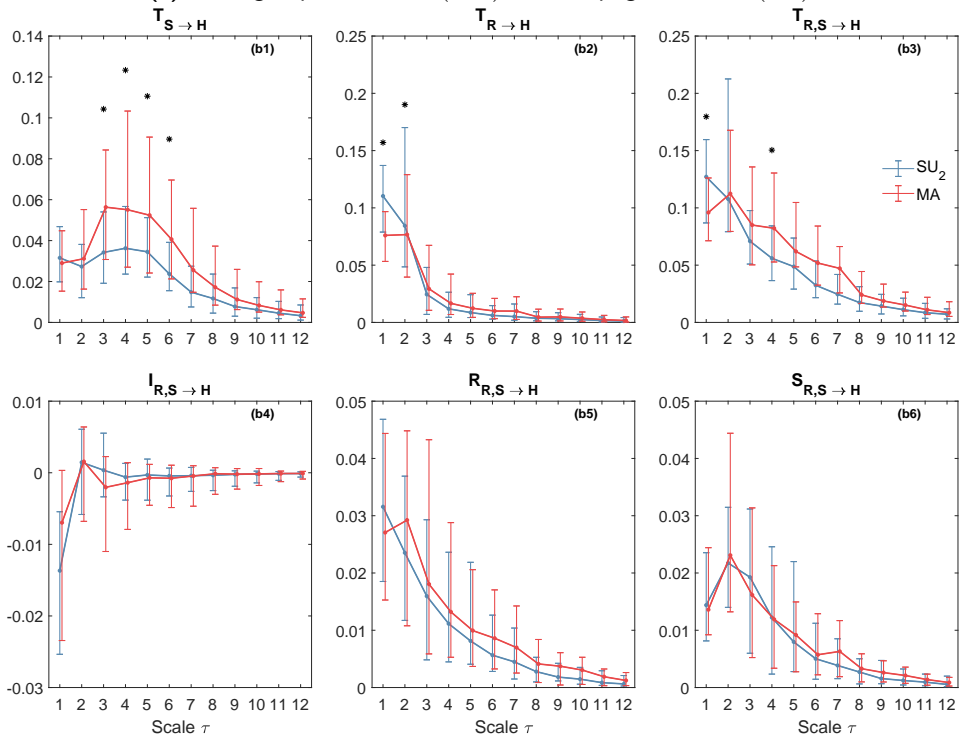}
   \caption{Recovery Supine Position $(\mathrm{SU}_2)$ \textit{versus} Mental Arithmetic's $(\mathrm{MA})$.}
   \label{fig:Analysis_b}
\end{subfigure}

\caption{Median and quartiles of the information measures across subjects during: \textbf{(a)} resting supine $\left(\mathrm{SU}_{1}\right)$ and postural stress $\left(\mathrm{UP}\right)$; \textbf{(b)} the recovery supine $\left(\mathrm{SU}_{2}\right)$ and mental stress $\left(\mathrm{MA}\right)$.}
\label{fig:Interquartiles}
\end{figure*}

Figure \ref{fig:Interquartiles} presents the median and quartiles across subjects of the six information measures computed as a function of the time scale $\tau=1, \ldots, 30$, for $\mathrm{SU}_{1}$ vs. $\mathrm{UP},$ panel \textbf{(a)} and $\mathrm{SU}_{2}$ vs. $\mathrm{MA},$ panel \textbf{(b)}. Statistically significant changes $(p<0.05)$ in TE measures at each time scale across the pairs $\mathrm{SU}_{1}$ vs. $\mathrm{UP}$ or $\mathrm{SU}_{2}$ vs. $\mathrm{MA}$ are marked with $*$. From a visual inspection of the multiscale patterns one can infer a markedly higher $T_{S \rightarrow H}$ at lower scales up to $\tau \approx 10$ moving from $\mathrm{SU}_{1}$ to $\mathrm{UP},$ panel \textbf{(a)}. The values of $T_{R \rightarrow H}$ are lower in UP for scale 1. In contrast, for scales 3 – 7 this measure is higher in UP phase. We can observe a similar behaviour in the $T_{S,R \rightarrow H}$ where the values of this measure are significantly higher from $\tau=2$ to $\tau=9.$ The $I_{S,R \rightarrow H}$ decreases significantly with tilt, mainly in the first three time scales. Then, significant differences between $\mathrm{SU}_1$ and $\mathrm{UP}$ can be observed in the mid-range time scales $\tau=5,6,7$. Observing $R_{S, R \rightarrow H}$ and $S_{S, R \rightarrow H}$ as two distinct information modification measures we see that both exhibit significant differences up to time scale 7 with higher values in the upright phase ($\mathrm{UP}$). Note that in the first few time scales the redundancy is greater than the synergy, hence the negative values of ITE in these time scales.

Moving from $\mathrm{SU}_{2}$ to $\mathrm{MA}$, we find significantly higher values of $T_{S \rightarrow H}$ in the mid-range of time scales. The $T_{R \rightarrow H}$ exhibits statistically significant differences (a decrease) only for the first scales. Similar behaviour is observed for the joint TE, with significant higher values in the $\mathrm{MA}$ phase for $\tau=1,4$. Regarding $I_{S,R \rightarrow H}$, $R_{S,R \rightarrow H}$ and $S_{S,R \rightarrow H}$ the model was not able to find significant differences. The profile of  $I_{S,R \rightarrow H}$, $R_{S,R \rightarrow H}$ and $S_{S,R \rightarrow H}$ in recovery supine position ($\mathrm{SU}_2$) and in the mental stress phase ($\mathrm{MA}$) are quite similar. For this reason, no significant difference were detected.

\begin{figure*}[ptbh]
\centering
\subfloat[Resting Supine Position ($\mathrm{SU}_1$) \label{fig:EMM_SU1}]
        {\includegraphics[width=0.49\textwidth]{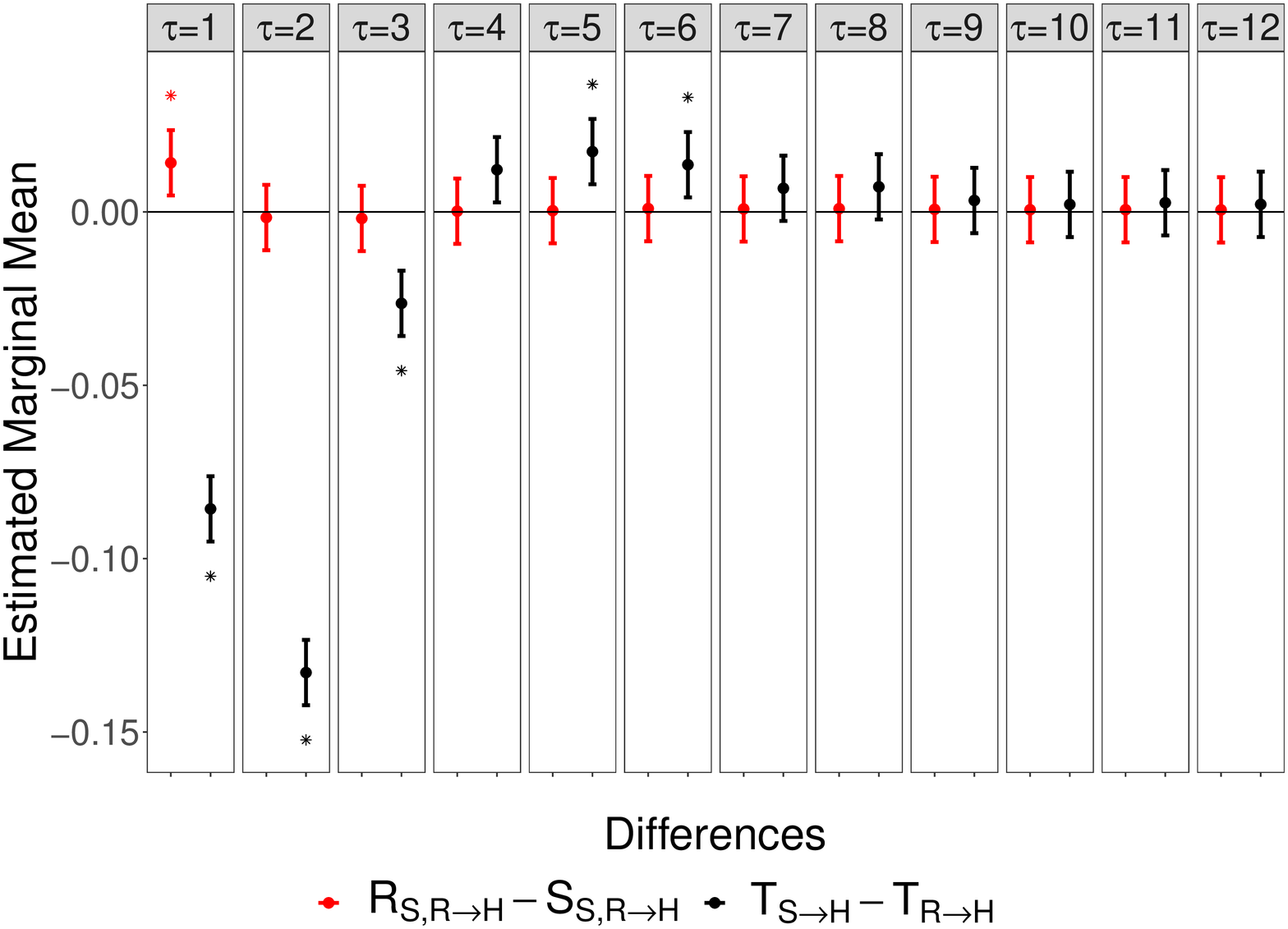}}
    \hfill
\subfloat[Upright Position ($\mathrm{UP}$) \label{fig:EMM_UP}]
         {\includegraphics[width=0.49\textwidth]{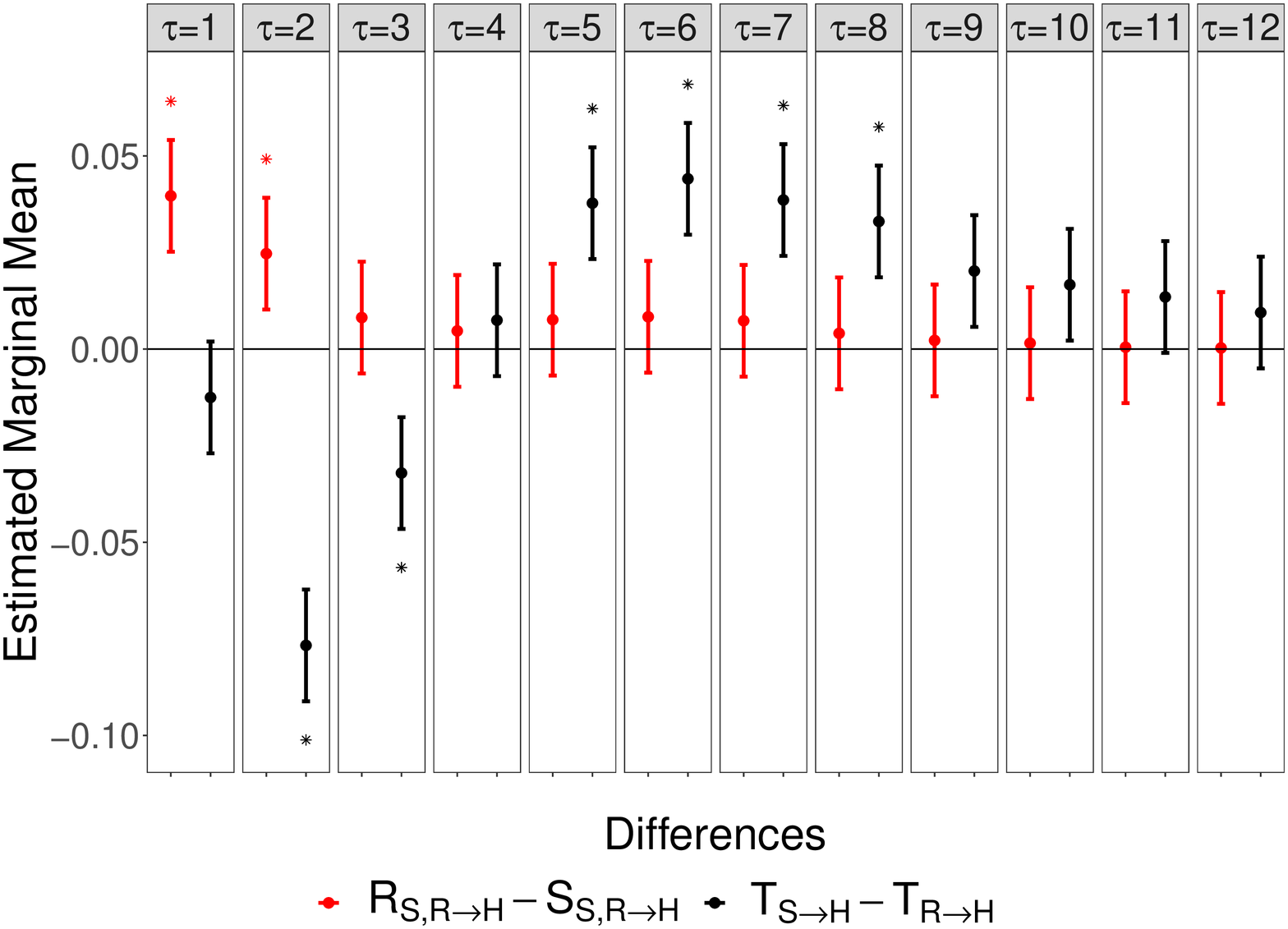}}

\subfloat[Recovery of Supine Position ($\mathrm{SU}_2$) \label{fig:EMM_SU2}]
         {\includegraphics[width=0.49\textwidth]{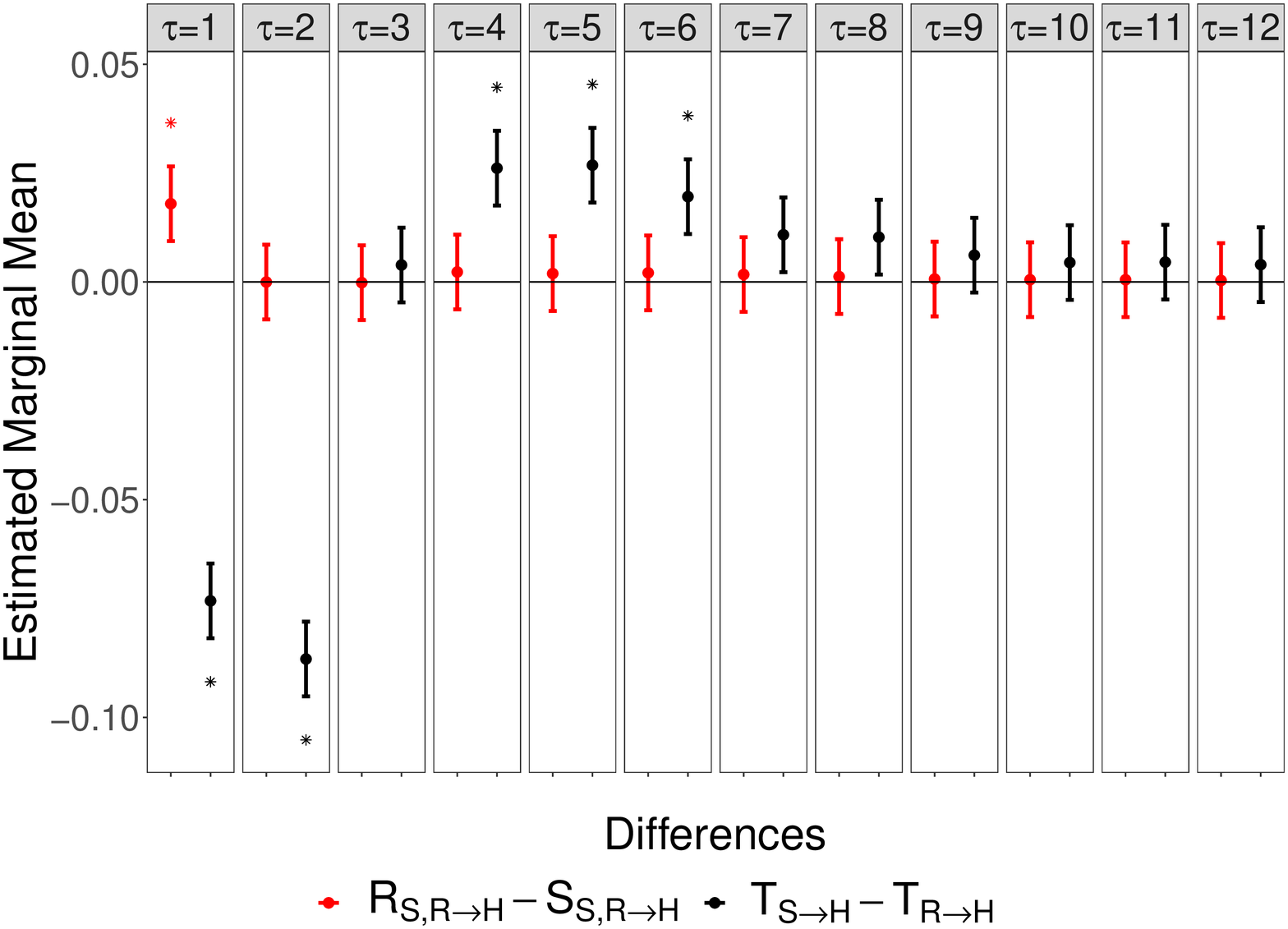}}
    \hfill
\subfloat[Mental Arithmetrics ($\mathrm{MA}$) \label{fig:EMM_MA}]
        {\includegraphics[width=0.49\textwidth]{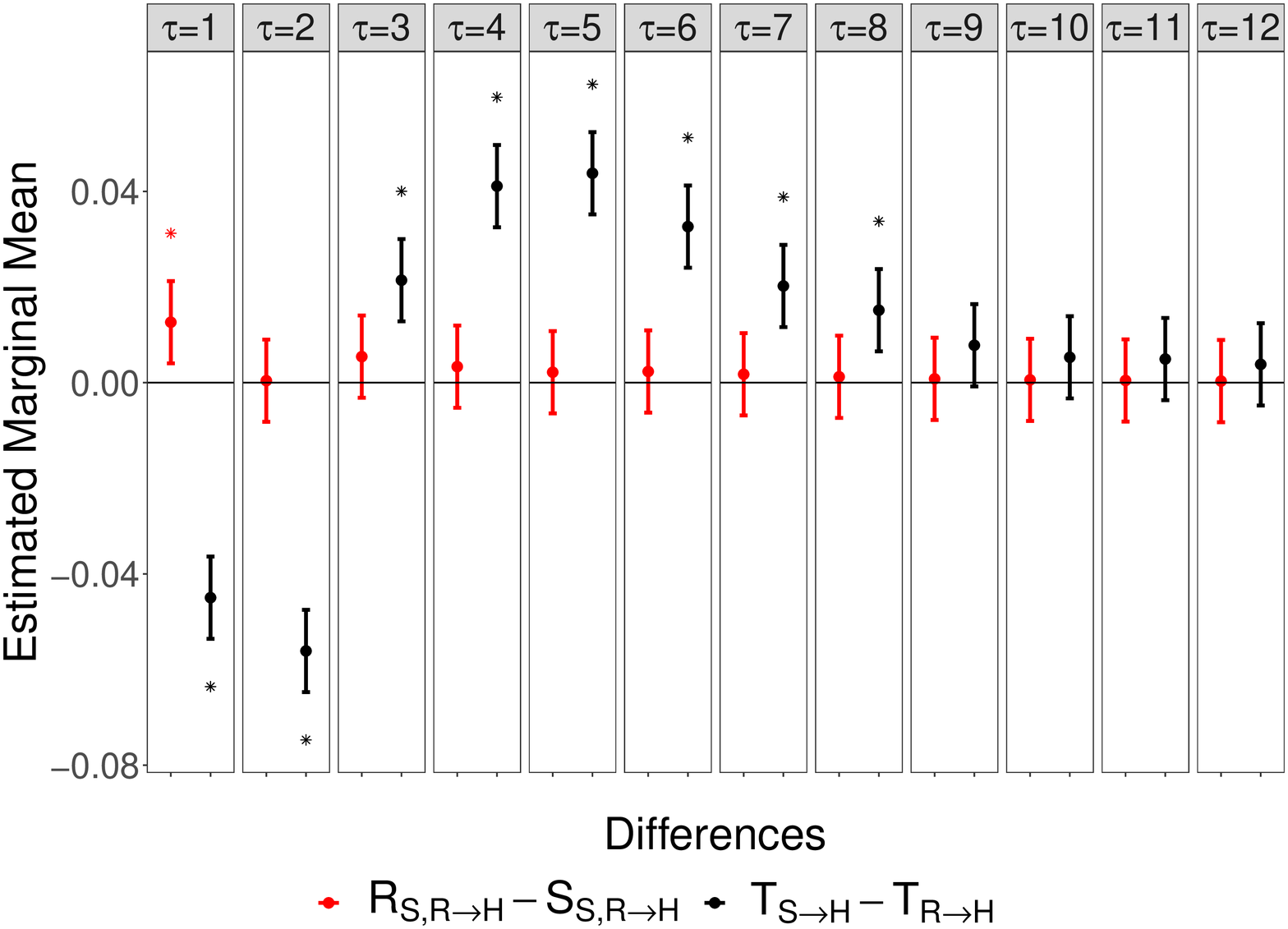}}
\caption{Estimated Marginal Means for the differences of $T_{S \rightarrow H}-T_{R \rightarrow H}$ (black) and $R_{S,R \rightarrow H}-S_{S,R \rightarrow H}$ (red) for the 4 phases of the experimental protocol.}
\label{fig:EMM Measures}
\end{figure*}

\Fref{fig:EMM Measures} presents the estimated marginal means for the differences between $T_{S \rightarrow H}-T_{R \rightarrow H}$ and $R_{S,R \rightarrow H}-S_{S,R \rightarrow H}$  and the respective 95\% confidence intervals for the 4 positions of the experimental protocol. Statistically significant differences $(p<0.05)$ are marked with $*$.

In the resting supine position ($\mathrm{SU}_1$, Fig.\ref{fig:EMM_SU1}) we find prevalence of the redundancy over synergy at $\tau=1$. For the other time scales, no significant differences are observed for these two measures of information modification. Comparing $T_{S \rightarrow H}$ and $T_{R \rightarrow H}$, the $T_{R \rightarrow H}$ overcomes the $T_{S \rightarrow H}$ in the first three time scales. This behavior is inverted at $\tau=5,6$ where the value of $T_{S \rightarrow H}$ is superior. In the remaining time scales no significant differences were found.

Moving to the upright position ($\mathrm{UP}$, Fig.\ref{fig:EMM_UP}) a prevalence of the redundancy is again observed in the first time scales $\tau=1,2.$ For the other time scales no significant differences between $R_{S,R \rightarrow H}$ and $S_{S,R \rightarrow H}$ were detected. Regarding the individuals TEs $T_{S \rightarrow H}$ and $T_{R \rightarrow H}$ we observed a similar behaviour to that observed in the $\mathrm{SU}_1$, that is, the oscillations observed in the heart period are essentially of respiratory origin. For the mid-range scales $\tau=5,6,7,8$, an inversion is observed, where oscillations of vascular origin prevail over those of respiratory origin. No significant differences were found for the remaining time scales.

The results of the estimated EMM for the differences in recovery supine position ($\mathrm{SU}_2$, Fig.\ref{fig:EMM_SU2}) are similar to those observed previously. We find only one significant positive difference for $R_{S,R \rightarrow H}-S_{S,R \rightarrow H}$ at $\tau=1$ denoting a preponderance of redundancy. Regarding $T_{S \rightarrow H}-T_{R \rightarrow H}$, we note prevalence of the $T_{R \rightarrow H}$ at $\tau=1,2.$ On the other hand, at $\tau=4,5,6$ the individual TE $T_{S \rightarrow H}$ plays a more dominant role.

Finally, in the mental stress phase ($\mathrm{MA}$, Fig.\ref{fig:EMM_MA}) a prevalence of the redundancy is observed at the first time scale and no significant difference $R_{S,R \rightarrow H}-S_{S,R \rightarrow H}$ was found in the remaining scales. The $T_{S \rightarrow H}-T_{R \rightarrow H}$ are significant up to time scale $\tau=9$, however in the first two temporal scales $\tau=1,2$ the oscillations observed in the heart period are predominantly of respiratory origin. In the remaining time scales, i.e., $\tau=3 \ldots 8$, oscillations of vasomotor origin have a leading role.

\section{Discussion} \label{sec:Discussion}

The present study extends multiscale information decomposition to analyze interactions inside cardiovascular and respiratory control systems accounting for the presence of short term dynamics and long-range correlations. The proposed linear parametric framework retains the advantage of previous formulations \citep{faes2017multiscaleGranger,faes2017multiscale} and incorporates long-range dynamics, which is fundamental for proper evaluation of information transfer at coarse time scales.
 
The increase of transfer entropy from S to H at lower time scales (\Fref{fig:Interquartiles} a1) when going from rest to tilt is in agreement with previous studies \citep{faes2011information,krohova2019multiscale,westerhof2006time} indicating indicating the dominance of baroreflex-mediated interactions during orthostatic
stress. The statistically significant differences evidenced for mid-range time scales also indicate that S has the most relevant part of its dynamics in the LF and VLF bands, as already noted in \citep{krohova2019multiscale}, and that postural stress induces changes to such dynamics noticeable in a wider range of time scales. For the $\mathrm{UP}$ position, at $\tau = 1$ the information transfer from R to H (\Fref{fig:Interquartiles} a2) is significantly lower than in rest. This is in accordance with previous findings evidencing that the respiratory sinus arrhythmia (RSA) decreases with tilt \citep{javorka2018towards,krohova2019multiscale,porta2012model}, probably due to the decreased parasympathetically mediated RSA. The trend is different for longer time scales, at which the information transfer from R to H becomes significantly higher than in rest, indicating the prevalence of slower oscillations in the information transfer from R to H, i.e. slowly varying respiration influences (mostly related to spontaneous changes of the respiratory pattern) are transferred more to slower HRV oscillations during postural stress \citep{krohova2019multiscale}. The two previous effects produce a higher joint information transfer (\Fref{fig:Interquartiles} a3) during UP for time scales from 2 to 9,  denoting stronger redundancy, as seen in \Fref{fig:Interquartiles} a4 from the negative I and from \Fref{fig:Interquartiles} a5-a6 (higher redundancy than synergy) for time scales up to 7. Physiologically, such results confirm previous findings indicating that postural stress produces a strong activation of  baroreflex-mediated RSA, path $R \rightarrow S \rightarrow H$, especially for time scales longer than $\tau > 1$. This confirms that baroreflex is a fundamental mechanism for slower heart rate oscillations, which should be studied in the LF band \citep{krohova2019multiscale,pernice2021comparison}. 

Regarding the comparison between mental stress and $\mathrm{SU}_2$, the transfer entropy $T_{S \rightarrow H}$ is higher during $\mathrm{MA}$ if compared to rest at mid-range time scales ($3 \leq \tau \leq 6$) (\Fref{fig:Interquartiles} b1), reflecting the activation of vasomotion associated with enhancement of slower blood pressure oscillations. Conversely, there is a significant decrease of information transfer from R to H at lower time scales  ($\tau=1,2$) during mental stress (\Fref{fig:Interquartiles} b2), indicating an overall weakening of the influence of respiration on heart rate, which is in agreement with vagal inhibition, but also the lower involvement of baroreflex-mediated RSA, provoked by stress challenges already demonstrated in previous works \citep{javorka2018towards,krohova2019multiscale,faes2011information}, and it is also in accordance with reduced cardiorespiratory interactions and synchronization occurring during mental task \citep{pernice2020multivariate,widjaja2013cardiorespiratory,zhang2010effects}. The combination of such results highlights that mental stress not only produces the weakening of RSA due to vagal inhibition, but also the nonactivation of baroreflex-mediated RSA ($R \rightarrow S \rightarrow H$), differently from postural stress \citep{krohova2019multiscale}. The results of joint information transfer (\Fref{fig:Interquartiles} b3) evidence the progressive increase occurring during the mental stress from time scale 1 to time scale 4: for $\tau=1$ there is a significant decrease during $\mathrm{MA}$ driven by the prevalence of respiratory dynamics, then there is the activation of baroreflex effects for mid-range time scales that produces the prevalence of information transfer from S dynamics with regard to slower oscillations (the increase becomes statistically significant in $\mathrm{MA}$ just for $\tau=4$). Such effect is not statistically significant for $\tau=5$ and higher scales, differently from postural stress (\Fref{fig:Interquartiles} a3), given the lower influence of respiratory dynamics already starting from $\tau=3.$ This may be due to the the combined effect of (1) vasomotor reactions elicited by MA through cortical mechanisms reflected in SAP changes and then transferred to HRV through the baroreflex and (2) the reduced RSA due to the significant increase in the breathing rate during $\mathrm{MA}$ if compared to $\mathrm{SU}_2$ (in agreement with \citep{krohova2019multiscale}). Such a difference between the stress condition and the corresponding rest status among the various time scales emphasizes the importance of employing a multiscale approach when studying cardiovascular and cardiorespiratory interactions, suggesting also that such interactions include complex multiscale patterns which respond flexibly to stress challenges \citep{krohova2019multiscale,widjaja2013cardiorespiratory}. No statistically significant differences between $\mathrm{MA}$ and $\mathrm{SU}_2$ are detected, at any time scale, analyzing the interaction transfer entropy (\Fref{fig:Interquartiles} b4), and neither with regard to redundancy (\Fref{fig:Interquartiles} b5) or synergy (\Fref{fig:Interquartiles} b6), conversely to postural stress, evidencing that the mechanisms underlying postural and mental stress are different, not only when considered "raw" (i.e. at $\tau=1$) but when going thoroughly a multiscale analysis. This also supports the importance of employing a multiscale approach to shed more light on such mechanisms, and its potential usefulness to differentiate between stress conditions.

The analysis of marginal means (\Fref{fig:EMM Measures}) indicates the prevalence of redundancy only for $\tau=1$ during $\mathrm{SU}_1$, and also for $\tau=2$ during tilt, and this can be put in relation to the activation of baroreflex due to the postural stress. Moreover, the difference between $T_{S \rightarrow H}$ and $T_{R \rightarrow H}$ indicates for $\tau<4$ the prevalence of information transfer from respiration to heart rate, evidencing the dominance of RSA on short time scales. The opposite is instead observed for mid-range scales ($\tau$ from 5 to 9), thus confirming once again that S has the most relevant part of its dynamics in the LF and VLF bands, and that transfers more information to H than respiration when assessed only for slower oscillations \citep{krohova2019multiscale}. Similar trends are reported comparing the two different stress typologies ($\mathrm{MA}$ and $\mathrm{UP}$), with the only difference being evidenced for $\tau=3$, with a prevalence of redundancy for S at $\tau=2$ during postural stress. Conversely, we notice an increase of information transfer from S (instead than R) to H during mental stress, evidencing that mental stress strengthens respiration-unrelated baroreflex effects. Overall, the obtained results highlight that head-up tilt induces scale-dependent variations in the transfer entropy of arterial pressure, higher in the mid-scales associated with slow oscillations, and lower associated to the effects of respiration. This result is similar to what observed in \citep{faes2019multiscalePRE} with regard to complexity of arterial pressure time series.

\section{Final Remarks} \label{sec:Final}
The aim of this study was to introduce an analytical framework, for multivariate Gaussian processes, where both IID and PID can be exactly evaluated in a multiscale fashion. Due to its parametric formulation, the method presented in this work inherits the computational efficiency of linear multiscale entropy \citep{faes2017efficient}, and more importantly, the VARFI modeling allows the description of both the short term dynamics and the long term correlations. Since long-range correlations are a crucial component of multiscale dynamics, this approach opens the way for a reliable estimation of the information modification of a variety of natural and man-made processes in which distinct mechanisms coexist and operate across multiple temporal scales. The application of this approach to the cardiovascular and cardiorespiratory dynamics highlights the scale-dependent variations of the information transfer measures of the signals herein considered.

Future developments of this work encompass the refinement of the SS model structure to support the description of long-range correlations \citep{sela2009computationally} with possible extension to nonstationary and cointegrated processes \citep{kitagawa1987non,Johansen2019, Gil-Alana2020}.
The applicability of this method of analysis to non-Gaussian processes, with accurate analytical solutions or computationally-reliable estimating methodologies, remains a fundamental challenge in the field. This is an important direction for future research since real-world processes frequently feature non-Gaussian distributions.

In terms of application contexts, the methodology proposed in this study can be exploited to characterize the altered cardiovascular and respiratory dynamics in a range of pathological states, e.g. including diabetis \citep{Sorelli} and pulmonary fibrosis \citep{Santiago-Fuentes}.
Moreover, the study of multivariate multiscale dynamics is particularly interesting in econometrics \citep{zhang2019multivariate} and neuroscience \citep{courtiol2016multiscale}, where dynamics spanning many temporal scales are frequently observed and multichannel data gathering is widespread. In particular, the methodology proposed in this work can be very useful to study the effects in longer time scales of the interactions between brain and the heart \citep{pernice2021multivariate,silvani2016brain,almeida2017arfima}.

\ack
This work was partially supported by  UIDB/00144/2020 (CMUP) and LA/P/0063/2020 (LIAAD-INESC TEC), which are financed by national funds through FCT – Funda\c{c}\~{a}o para a Ci\^{e}ncia e a Tecnologia, I.P. . Riccardo Pernice is supported by the Italian MIUR PON R\& I 2014-2020 AIM project no. AIM1851228-2. Luca Faes is supported by the Italian MIUR PRIN 2017 project 2017WZFTZP “Stochastic forecasting in complex systems”. Michal Javorka is supported by grants VEGA 1/0199/19, 1/0200/19 and 1/0283/21.

\bibliographystyle{agsm}
\renewcommand\harvardurl{\textbf{URL:} \url}
\bibliography{MyBib}

\end{document}